# Structural properties of liquid cesium metal by applying cohesive energy density of the liquid state

M. H. Ghatee[*], M. Sanchooli[**]

(*Department of Chemistry, Shiraz University, Shiraz, Iran 71454*)

Fax: +98 711 228 6008
E-mail: ghatee@susc.ac.ir

___

[*]Corresponding Author, [**]Permanent address: Department of Chemistry, Zabol University, Zabol, Iran,

# Structural properties of liquid cesium metal by applying cohesive energy density of the liquid state

Abstract

The structural property of liquid cesium is investigated in the temperature range 900 K to 1900 K by application of semiempirical effective Lennard-Jones (8.5-4) pair potential function and employing Gillan's algorithm to solve Percus-Yevick equation. The potential function has been derived accurately by application of cohesive energy density in a wide range of pressure-density-temperature ($P\rho T$) data including data at the proximity of absolute zero temperature. The method is very much responsive to the liquid dynamics and leads to indication of three distinct ranges of metal, nonmetal, and metal-nonmetal transition states. The resulted pair correlation functions are well compared with the reported experimental and first-principle molecular dynamic. The calculated coordination numbers in the liquid range are in agreement with experiment particularly at low temperatures, though it is singular at about 1400 K. This observation is similar to the change from one liquid structure to another one and is verified by heights of the first peak of experimental pair correlation function as a function of temperatures. At $T = 973\,\text{K}$, the position of the first peak ($r = 5.27\,\dot{\text{A}}$) is in agreement well with experimental ($r = 5.31\,\dot{\text{A}}$) and with the first-principle DFT molecular dynamics ($r = 5.24\,\dot{\text{A}}$).

## 1. Introduction

A great deal of attempt has been devoted to the theoretical and experimental study of expanded fluid alkali metals [1-13]. Alkalis, as elemental monovalent metals, closely resemble the expanded crystals with half full bands [14-15]. Most of these efforts have focused on cesium due to its relatively low critical temperatures. Studies, such as electrical conductivity and the equation of state [4], clearly show that near the critical point cesium undergoes a metal-nonmetal transition. Near the critical point the conductivity drops sharply, thus showing a strong effect of the phase transition on the electronic structure. However, except across the liquid-vapor phase boundary, no indication of a sharp (first order) electronic transition can be seen. Strong correlation between the behavior of the density and that of the electrical conductivity indicates that the variation of density is the dominant factor governing the metal-nonmetal phenomena. As the liquid cesium is expanded, the average coordination number decreases and average nearest-neighbor distance increases slightly. So in order to describe the properties of the liquid cesium as a function of density these factors have to be taken into account.

The study on the structure of liquid metal has been an important subject in the investigation of liquid properties. In this regard, the accuracy and the role of different parts of the pair potential function have been extensively studied for normal liquids (e.g., nonmetal, nonquantum fluids) [16]. The results indicate that at high densities the pair correlation function $g(r)$ is affected mainly by the short range repulsive part of the pair potential and that the long range attractive part only affects the height of the first peak [17]. The results, however, obtained for liquid Cs metal have been in sharp contrast with normal fluid [17]. Typically, at high density of liquid Cs (e.g., $T = 323\,\text{K}$, $\rho = 13767.23\,\text{mol/m}^3$), the long range attraction part has the effect of decreasing the height of first peak of the pair correlation function and conversely it has the effect of increasing the height at high temperature (e.g., $T = 1373\,\text{K}$, $\rho = 9107.97\,\text{mol/m}^3$).

In our previous studies, the prediction of equilibrium thermodynamic properties of liquid cesium by applying accurate potential functions was presented [18-20]. In particular we have presented an equation of state and evaluated the equilibrium thermodynamic properties of the liquid state. Also transport properties of vapor state of cesium metal was predicted by applying accurate potential function [see eq. (1)] obtained from liquid cohesive energy density data including data for super-cooled state at the proximity of absolute zero temperature [21,22]. Kozhevnikov et al. [22], have used their experimental $P\rho T$ data of liquid cesium in the range 400 K-2000 K and have reported the internal pressure $P_{int}$. Using these data augmented with $P_{int}$ at the proximity of absolute zero of super-cooled state [23], then they have derived an effective pair potential. Thus, this potential function includes boundary condition $P_{int} \cong 0$ at $r = \sigma$ through $r = r_{min}$ and beyond, where $\sigma$ is the hard-sphere diameter and $r_{min}$ is the position of potential minimum. The boundary condition for $T$ belonging to low temperatures (e.g. super-cooled state) exactly matches the available experimental volume of cesium at $T = 5$. By their model, Lennard-Jones (8.5-4) potential function [22]

$$u(r) = A\varepsilon \left[ \left(\frac{\sigma}{r}\right)^{8.5} - \left(\frac{\sigma}{r}\right)^{4.0} \right] \tag{1}$$

where $u(r)$ is the effective pair interatomic potential, $\varepsilon$ is the potential well-depth, $r$ is the interatomic distance, $A = 3.6914$, the hard-sphere diameter $\sigma = (2.125)^{-1/4.5} r_{min}$. Although Eq. (1) is a model potential but it represents the force of cohesion of the molecules analytically in terms of molecular parameters.

In the pervious work, the two-parameter LJ (8.5-4) potential function has been used to predict equilibrium thermodynamic properties of liquid cesium metal [21]. To achieve this, it has assumed that the pressure exerted by a single molecule is composed in part due to the external pressure $P$, and due to the internal pressure $(\partial E/\partial V)_T$, which represent the cohesive forces of molecules, where $E$ is the internal energy of the system. The sum, which gives the thermal pressure $T(\partial P/\partial T)_V$, is the so called thermodynamic equation of state. Using Eq. (1) the thermodynamic equation of state is solved for $P$. The resulted isotherms allow calculation of molecular parameters of the potential function $r_{min}$ and $\varepsilon$ by using $P\rho T$ data of liquid state. As a result, $r_{min}$ and $\varepsilon$ are functions of temperature and thus $u(r)$ turns to be a temperature dependent interatomic potential function.

Also the equation (1) has been used to calculate the non-equilibrium properties (viscosity and thermal conductivity) of cesium vapor by using the method of Chapman–Enskog solutions of Boltzman equation. The range of accuracies of the results indicates that eq. (1) is quite accurate in describing interatomic interaction in liquid cesium [21].

In this work we investigate to present the structural properties, namely pair correlation function and structure factor, of cesium metal in the temperature range $900\,\text{K} - 1900\,\text{K}$. It is motivated by the fact that LJ (8.5-4)-potential function is accurately applicable to investigate thermodynamic equilibrium properties of liquid and transport properties of vapor states of cesium metal. The Gillan's algorithm for the numerical solution of Percus-Yevik (PY) equation is used.

The results are compared with the experimental $g(r)$, and with the structure factor $S(Q)$. Also comparisons are made with the results of first principle molecular dynamics (MD) in which density functional theory for electronic calculation is applied. The onset of and the properties of metal-nonmetal transition can be characterized by a rather sharp change in the height of $g(r)$. New results on the trend of pair correlation function which may be used to interpret metal-nonmetal transition are presented. Trend of the coordination number in relation to the liquid structure is reported in detail.

## 2. Pair correlation function and structure factor

Structural properties of the liquid state are vastly determined by employing PY equation [24]. In the integral equation theory, the Ornestein-Zernike relation is given by

$$h(r) - c(r) = \rho_n \int c|r - r'|h(r')dr \qquad (2)$$

where $h(r)$ is the total correlation function, $c(r)$ is the direct correlation function, and $\rho_n$ is the number density. The equation (2) leads a supplementary closure for the pair correlation function. The closure equation for the PY approximation is

$$g(r) = \exp(-\beta u(r))[1 + h(r) - c(r)] \qquad (3)$$

where $\beta = 1/kT$, $k$ is the Boltzman constant, and $T$ is the absolute temperature. Except for the computer simulation approaches, most methods for calculation of pair correlation function are based on the solution of integral equations. Gillan has presented a powerful method of obtaining numerical solutions to PY [and hypernetted chain (HNC) as well] for liquid structure [25]. For a system with particular force field, the inputs are reduced temperature, $T^* = kT/\varepsilon$, and reduced density, $\rho^* = \rho\sigma^3$, and output is the pair correlation function. The method is quite accurate and sensitive to the detail of potential function. It has been used successfully to investigate the role of the tail of potential function on the pair correlation function near triple point of liquid cesium [17].

In this study, LJ (8.5-4)-potential function is used to solve the PY integral equation in the temperature range from about the normal boiling point to the near critical point (e.g., $900\,\mathrm{K} - 1900\,\mathrm{K}$) of the liquid Cs. Calculations are made for liquid density at the saturation pressure. Values of $\varepsilon$ and $\sigma$ as a function of temperature is determine by the previous method [21]. Pair correlation functions versus reduced intermolecular distance $r^* = (r/\sigma)$, in the above mentioned temperature range are shown in Figure 1. As the temperature is increased the positions of the first peak are shifted slightly to the higher distance. Also the height of the first peak is reduced as the temperature is increased. This shows a clear indication of the dependence of $g(r)$ on the liquid density as well.

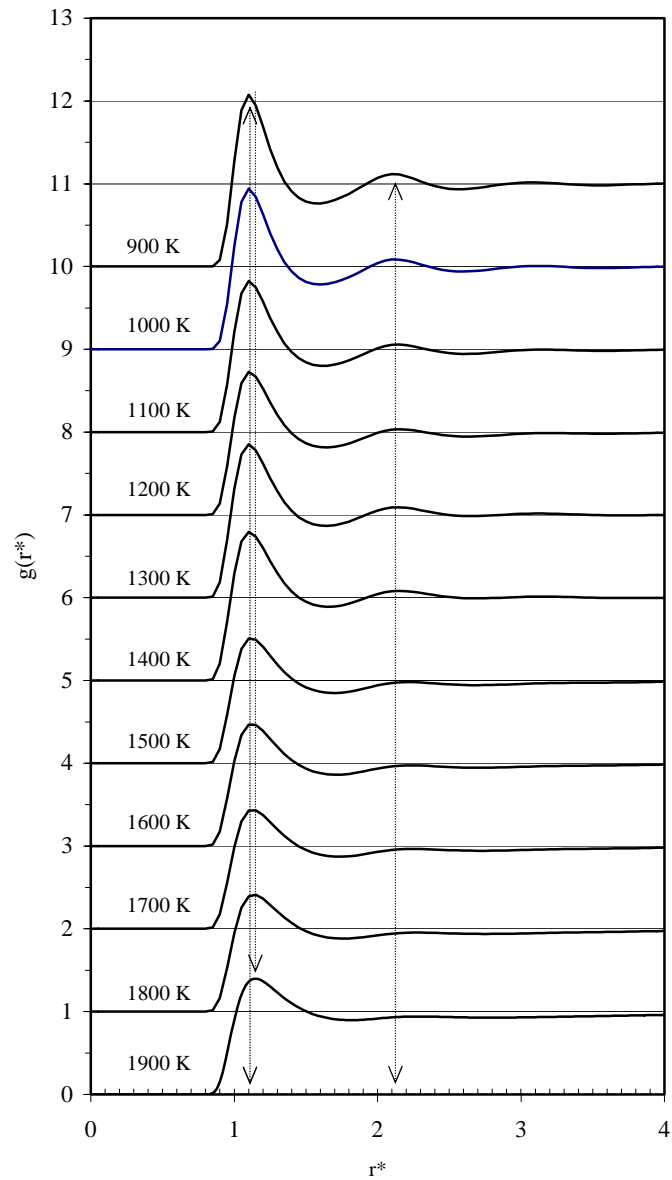

Figure 1. Pair correlation functions at saturation densities

Pair correlation functions thus obtained are transformed to the corresponding structure factors. Structure factors at 973 K, 1173 K, 1373 K, and 1673 K at saturation pressure have plotted in the Figure 2. Also the experimental structure factors extracted from reference 1 have plotted for comparison. Notice that the available experimental structure factors are at pressures higher than the corresponding saturation pressures except at 1173 K. Based on LJ (8.5-4) potential function, the presence of a relatively strong second peak corresponding to the influence of central atom on the second neighboring shell is very indicative in all cases.

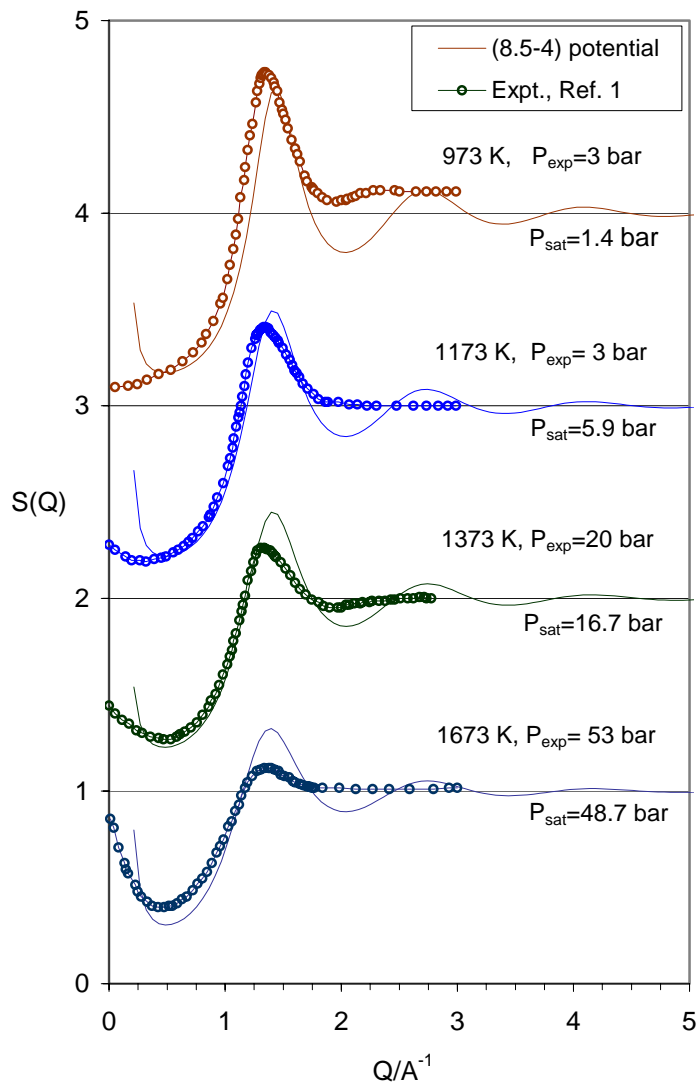

Figure 2. Structure factors of liquid cesium: calculated at saturation densities; experimental at specified pressure.

To study the structural properties of liquid cesium rigorously, Cabral et al [26], have used the density-functional-theory for electronic calculation to determine atomic force. The energy was represented by the local density Khon-Sham energy functional in the plane-wave pseudopotential formulism. They have generated then the dynamics of the system by integration of the equation of motion through the algorithms used in the classical MD simulation. In Figure 3, the resulted $g(r)$ at 973 K and saturation density of this study has compared with the first principle molecular dynamics [26], and with the experiment [27].

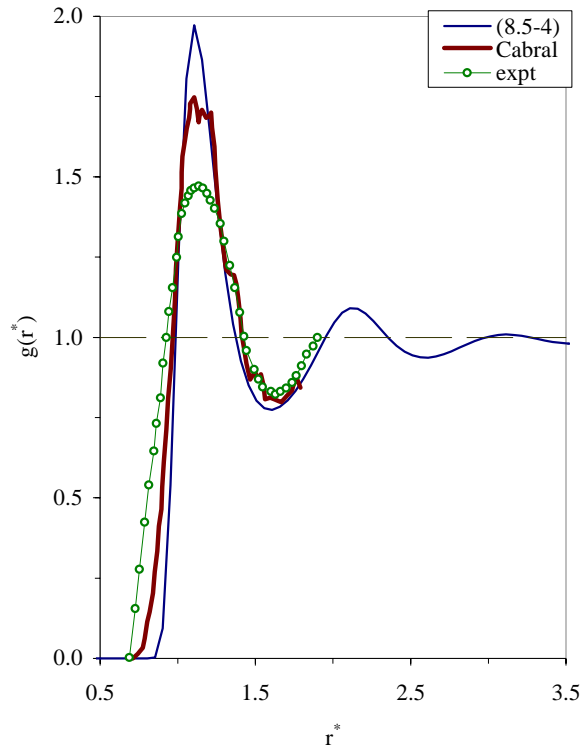

Figure 3. Comparison of $g(r)$'s at 973 K and saturation density: this work, the first-principle DFT Molecular Dynamic [26], and the experiment [27].

## 3. Results and Discussion

In the Gillan's method, the traditional method of Picard and Newton-Raphson technique has been combined to improve the instability of the former and slowness of the latter. The Numerical test for a typical potential function has shown that the method is well convergent within 20-30 iterations. The number of iterations needed is insensitive to the choice of initial estimate, however, it is applied only to a certain range of temperature and density. The drawback with the Gillan's is no convergence at real high density and low temperatures. On the other hand, because experimental $P\rho T$ data of a particular fluid system can be modeled by different potential function (leading to characteristics $\varepsilon$ and $\sigma$), the failure or success of Gillan's algorithm for convergence at a particular $T^*$ and $\rho^*$ depends on the applied model potential function. Fortunately we could get a good convergence for LJ (8.5-4) potential function within the wide range of temperature $900\,\text{K} - 1900\,\text{K}$ for densities at saturation pressures. No convergence could be obtained at $T^*$ and $\rho^*$ corresponding to $T < 900\,K$.

It has been shown that LJ (8.5-4) potential function can be used to predict equilibrium thermodynamic properties of liquid cesium as well as non-equilibrium thermodynamic properties of cesium vapor quite accurately [21]. The parameters of the potential function, from freezing temperature to near the critical temperature, are calculated by using the isotherm derived semiempirically and the application of $P\rho T$ data in the pressure range $50 - 600$ bar [21].

Pair correlation functions in the temperature range $900\,\text{K} - 1900\,\text{K}$ are shown overlaid in Figure 4. A close look at Figure 4 indicates that, based on the height of the first peaks and the behavior of the tails, there are three distinct temperature ranges with characteristics $g(r)$'s: (**a**) the range $900\,\text{K} - 1200\,\text{K}$, (**b**) the range $1200\,\text{K} - 1400\,\text{K}$, and (**c**) the range $1500\,\text{K} - 1900\,\text{K}$. [These ranges can be stated in terms of saturation densities too.] The trend of $g(r)$ in the range (**a**) is like that of monatomic and normal liquids that is, as temperature is increased, the height of first, second, and etc. are decreased steadily. On the other hand the tails oscillate more or less symmetrically and dump out at long intermolecular distances as in the case of monatomic and molecular normal fluids. In short it can be said that in range (**a**) the system is well-structured at short range as well as long range.

In range (**b**), the height of first peak abnormally increases as $T$ increases to $1300\,\text{K}$ and shifts back to lower value at $1400\,\text{K}$, yet it has a height larger than the height at $1200\,\text{K}$. The second peaks have appreciable heights whereas the tails dump out rather unsymmetrically at large $r$ with inappreciable heights. This can be used to represent the onset of metal-nonmetal transition as indicated by an inflection point on the responsiveness of the coordination number and a singular point in the trend of the height of the first peak of the experimental $g(r)$ (as will be discussed latter).

The most important feature of the present work is the behavior of the pair correlation function when one moves from temperature range (**b**) to the temperature range (**c**). One can see an enhancement of a phase transition on moving from (**b**) to (**c**). There is a sharp decrease in the height of the first peak as temperature is increased to $1500\,\text{K}$ and falls off steadily thereafter up to $1900\,\text{K}$. The trend in the range (**c**) has the characteristic of normal liquid, though the inherent nonmetal properties can not be deduced from these data. The density at saturation pressure is $8750.3\,\text{mol/m}^3$ at $1400\,\text{K}$ and $8163.4\,\text{mol/m}^3$ at $1500\,\text{K}$. Basically the density in range (**c**) decreases rather linearly with temperature, and thus it can be concluded that other factor such as binding energy of a molecule to other molecules is responsible for the structural behavior of the liquid. The first peak in this range is rather structured indicating that the first shell around a central atom is well structured even near critical point. Although it needs to be verified experimentally but this result can be justified by the fact that there is quite chemical effect in cluster formed by the effective association of few atoms, say 3 to 4 cesium atoms.

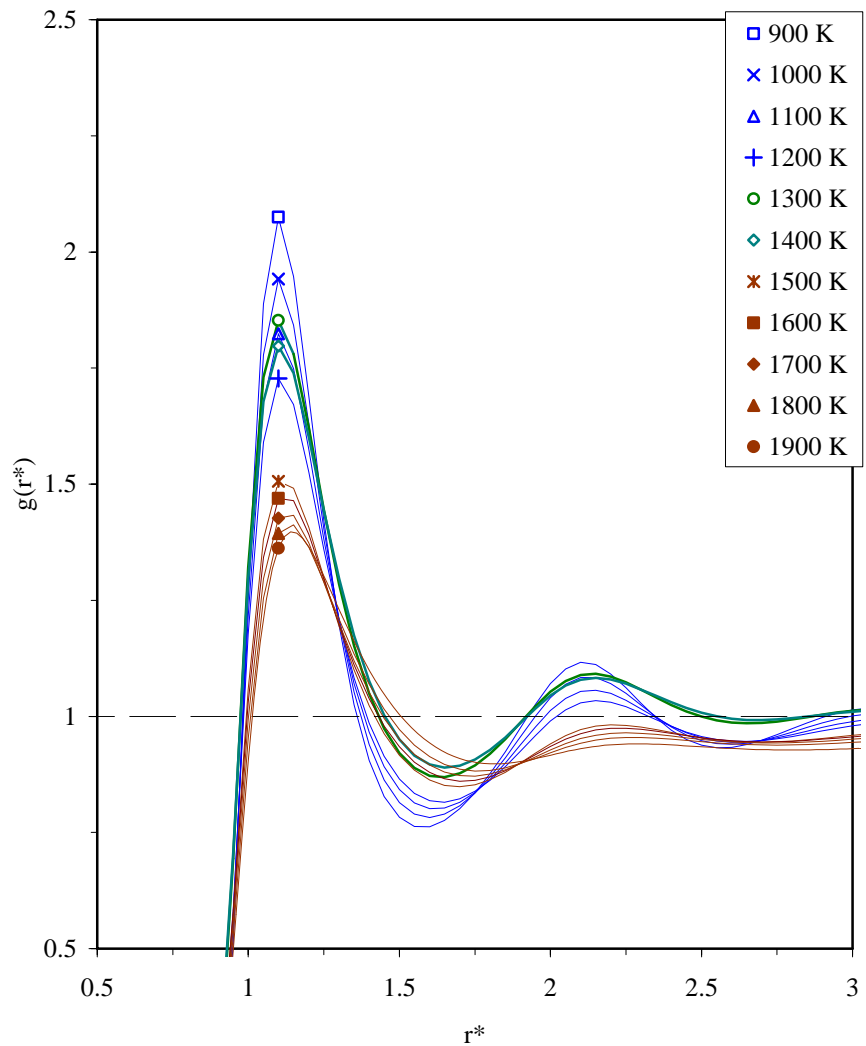

Figure 4. Comparison of the first peak and the tails of calculated pair correlation function of liquid cesium in the range $900\,\text{K} - 1900\,\text{K}$.

In the same range, the second peaks are totally unsymmetrical and structureless and tails stop oscillation, approaching barely to $g(r)=1$ from below at large $r$. This clearly indicates the absence of tightly bounded second shell and total absence of third shell. Therefore, a polyatom cluster, which is known to be formed in the case of alkali metal close to the critical temperature, might be assumed as a cluster of atoms comprising a single shell associated (rather chemically) with the central atom. We attribute this accurate prediction of metal-nonmetal transition to the fact that, (**i**) the LJ (8.5-4)-potential function has been obtained from experimental accurate $P\rho T$ data over a wide rang of pressure and temperature including data at the proximity of absolute zero temperature, and (**ii**) to fact that Gillan's method for solution of PY integral equation applies explicitly the molecular potential parameters and the experimental liquid density data (e.g., reduced temperature, $T^* = kT/\varepsilon$, and reduced density, $\rho^* = \rho\sigma^3$) as input. It should be recalled that LJ (8.5-4) potential function has been characterized by using experimental $P\rho T$ data and thus it is temperature dependent as well as density dependent.

The coordination number Z, at each temperature is determined by integration of the first peak of $g(r)$ [24]:

$$Z = 2\rho \int_0^{r_{max}} g(r)4\pi r^2 dr \tag{4}$$

where $r_{max}$ is the position of the first peak of the pair correlation function.

The Z values, obtained by applying the calculated $g(r)$'s to Eq. (4), are compared with the experimental value [28], in Figure 5. As it can be seen the coordination number changes rather linearly in the whole range. However, if one considers deviations from the linear behavior, an inflection point is seen clearly at about $1400\,\text{K}$. This inflection point is reported here for the first time. It needs to be verified experimentally, though, the experimental Z values are reported as a function of density in literature. On the other hand we have found experimental evidence corresponding to this point by considering the height of the first peaks of experimental $g(r)$ [27], as a function of temperature. As it can be seen from Figure 6, $g(r)$ tends to a singular behavior at about 1373 K.

Also our calculated structure factors for liquid cesium in the temperature range $973\,\text{K} - 1673\,\text{K}$ at saturation pressures are shown in Figure 2. In the same Figure, comparison is made with experiment at the (experimentally available) pressures [1], close to the corresponding saturation pressures. Although the experimental condition of the measured structure factors are not the same as the calculation, it can be seen that our result in general are in rather good agreement with experiment obtained by Winter et al. [1]. However, the height of the first peak of experimental structure factor decreases faster than the calculated one as the temperature is increased. On the other hand, the position of the first peak of calculated structure factor is smaller than the experiment at all temperature. This deviation can be attributed in part to the difference in pressure between experimental pressure and the saturation pressure.

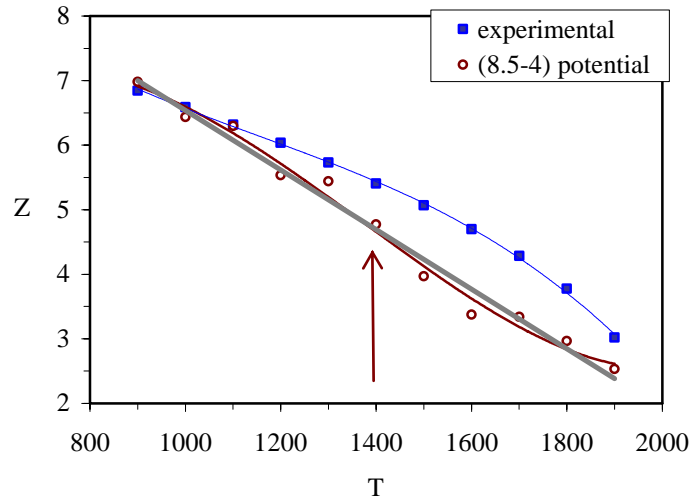

**Figure** 5. The Coordination number obtained by LJ (8.5-4) potential. (Linear gray line is the linear fit; reddish line is nonlinear fit.) The experimental results are shown for comparison. (Line is trend line.). The arrow shows the position of inflection point.

The reported $g(r)$ by the first-principle density functional MD are at the saturation density and at temperatures $350\,\text{K}$ (highest density), $573\,\text{K}$, and $973\,\text{K}$ (lowest available density). At $973\,\text{K}$, the result of this work is in agreement with the first principle MD over all the length scale considered as well as with the experiment (see Figure 3). However, the height of the first peak of our results lies above the first principle and the experiment. It can be seen that at $973\,\text{K}$ the position of the first peak of this study ($r = 5.27\text{Å}$) is in good agreement with first-principle ($r = 5.24\text{Å}$) and with the experimental result ($r = 5.31\text{Å}$). On the other hand, the value of distance of closest approach resulted by first principle ($r_c = 3.8\text{Å}$) lies between that of this work ($r_c = 4.3\text{Å}$) and that of experiment ($r_c = 3.3\text{Å}$). This might account for relative height of the first peak resulted by three methods. Although pseudopotential (used in the first principle MD method) is a soft potential as required by the nature of its construction, but it looks that it is still harder than the real system (in comparison with experiment). Another reason is the absence of a well-defined cut off distance usually assigned to identify core electron from orthogonal valence electron, which is an important task in construction of characteristic pseudopotential.

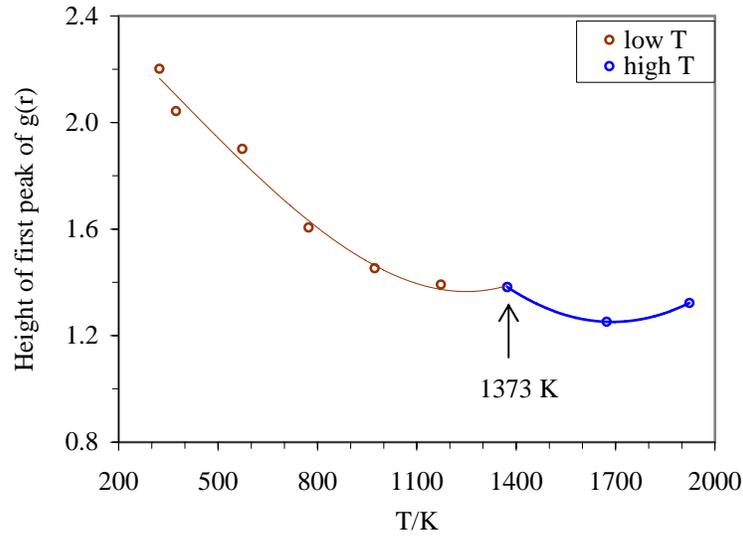

Figure 6. The height of the first peak of the pair correlation obtained by experimental method.

The structural properties of a dense fluid are determined by first-principle MD simulation based on the determination of atomic force using density-functional-theory calculation of the electronic structure. The most advantage of the first principle MD is that it avoids the difficult task of constructing an adequate potential function to model the interaction in condensed phases. The number of particle that can be used in the suppercell has limitation of first-principle MD simulation and only 54 Cs atoms have been involved [26]. This limitation of the first-principle MD is that with situation imposed by computational time resources the number of particles used in the simulation is limited and the dynamic time is last few picoseconds. The number of particles is small compared to classical molecular-dynamics simulation, limiting, for example the range of pair correlation functions studied (for liquid cesium metal) up to only $9\dot{A}$, the typical length scale of current first principle MD simulation. At this distances there are more tolerance (see Figure 3) reflecting the surface effect due to the limitation of the size of the supercell. However, application of an accurate known pair potential function has the advantage of avoiding such a limitation and the behavior of the tail can be monitored rather exactly.

As the final remarks, the position of the first peak of $g(r)$ obtained by application of (8.5-4) potential function is increased only slightly with temperature, in accord with the observation made by experimental results. On the other hand the coordination number as a function of temperature is in close agreement with the experiment and furthermore reveals a singularity characteristics of phase transition.

Gillan's methods did not converge for HNC for the system we considered in this study, and therefore we can not make comments about the relative accuracy of $g(r)$ obtained the solution of PY equation.

## Conclusion

This article provides new structural properties of liquid cesium metal using LJ (8.5-4) potential function obtained semiempirically by application of cohesive energy density of liquid state over a wide range of temperature including data at the proximity of absolute zero. Validity and range of accuracies of LJ (8.5-4) potential function have been already established by its application for equilibrium thermodynamic properties of liquid state and for the transport properties of the vapor state. Three distinct states are identified and attributed to the metal, metal-nonmetal transition, and nonmetal states. An abrupt decrease in the height of the first peak and a systematic behavior of the tail at 1500 K are clearly characterized and attributed to transition from metal to nonmetal state. The calculated pair correlation is in agreement with the experiment. Also, the calculated pair correlation function at 973 K is in agreement with the results of first-principle DFT molecular dynamic. Consistent with the experiment, it is found that the position of first peak increases slightly by about 6.4% from 900 K to 1900K, and the calculated coordination number changes rather linearly from about 7 at 900 K to 2.5 at 1900K. An inflection point in the trend of calculated coordination number and a singularity in the trend of the height of the first peak of experimental $g(r)$ are investigated and new information in the transition from a metal state to a nonmetal kind are undertaken.

**Acknowledgement**

The authors are indebted to the research council of Shiraz University for supporting this study. M. Sanchooli acknowledges the leave of absence granted by Zabol University and the **Ministry** of Science, Research and Technology for financial supports.

## List of symbols

| | |
|---|---|
| P | pressure |
| V | molar volume |
| T | absolute temperature |
| r | interatomic distance |
| g(r) | pair correlation function |
| h(r) | total correlation function |
| c(r) | direct correlation function |
| k | Boltzman constant |
| A | constant |
| E | total energy |
| S(Q) | structure factor |
| Z | coordination factor |

## Greek Symbols

| | |
|---|---|
| $\sigma$ | hard-sphere diameter |
| $\varepsilon$ | potential well-depth |
| $\rho$ | molar density |

## Subscript

| | |
|---|---|
| int | internal |
| min | minimum |
| n | number |
| max | maximum |

## superscript

| | |
|---|---|
| * | reduced |